# Highly Stable Silicon Anodes Enabled by Sub-10 nm Pores and Particles

*Pankaj Ghildiyal[1]\*, Brandon Wagner[2], Jianjun Chen[3], Tu Nguyen[3], Aishwarya Belamkar[4], Juchen Guo[2,3], and Lorenzo Mangolini[2,4]\**

1 SiLi-ion Inc., Multidisciplinary Research Building, 3401 Watkins Dr, Riverside, California 92507, United States

2 Materials Science and Engineering Program, University of California Riverside, Riverside, California 92521, United States

3 Department of Chemical and Environmental Engineering, University of California Riverside, Riverside, California 92521, United States

4 Department of Mechanical Engineering, University of California Riverside, Riverside, California 92521, United States

AUTHOR INFORMATION

**Corresponding Authors**

\*Email: pankaj@siliion.com

\*Email: lmangolini@engr.ucr.edu

ABSTRACT: Silicon anodes offer high energy densities for next-generation lithium-ion batteries; however, their application is limited by severe volume expansion during cycling. Making silicon porous or nanostructured mitigates this expansion but often increases lithium inventory losses due to the inherent high surface area of nanomaterials. This study introduces a simple bottom-up process that overcomes this limitation. The approach relies on small silicon particles (<10 nm) produced using an efficient low-temperature plasma approach. These small building blocks are assembled into micron-scale superstructures characterized by uniformly dispersed sub-10 nm pores. This structure addresses both volume expansion and lithium-inventory issues while achieving tap densities exceeding those of commercial graphite (~1.2 g·cm$^{-3}$), all while maintaining good processability. The resulting silicon-dominant anodes achieve remarkable stability in full pouch cells with NMC811 and LFP cathodes, retaining ~80% capacity for more than 400 cycles without pre-lithiation, graphite blending, or pre-cycling.

1. **Introduction**

Silicon is an attractive material for next-generation lithium-ion battery anodes, offering high gravimetric and volumetric storage capacity (3579 mAh g$^{-1}$ and 2194 mAh cm$^{-3}$) compared to graphite (350 mAh g$^{-1}$, 719 mAh cm$^{-3}$), as well as high earth abundance.[1] However, poor electrochemical stability has hindered the widespread adoption of silicon-dominant anodes, primarily due to the extreme volume change during cycling. The volume expansion results in a cascade of failure mechanisms, including mechanical failure, pulverization, and loss of electrical contact.[2,3] Researchers have extensively explored nanostructuring as a viable approach to mitigate these issues,[4–6] including strategies such as using nanowires or carefully engineering void spaces around silicon particles, among many others.[7,8] While these approaches effectively accommodate volume expansion and reduce mechanical stress in the anode, they introduce new complexities that limit performance and manufacturability.

A critical challenge in nanostructured silicon anodes lies in understanding and optimizing the interplay between particle size and porosity. The increased surface-to-volume ratio in these structures leads to heightened reactivity with the electrolyte, resulting in significant initial lithium-inventory losses. These issues are exacerbated by the large volume changes during cycling, which compromise the formation of a stable solid electrolyte interphase (SEI) and continuously expose fresh silicon surfaces to the electrolyte during cycling.[9] Pre-lithiation or pre-cycling methods have been proposed to address initial lithium inventory losses.[10,11] However, these approaches do not address the underlying issue of continuous SEI formation, and introduce additional manufacturing steps and barriers to commercialization. Furthermore, nanostructured materials face practical challenges in large-scale manufacturing. Lower tap density and high viscosity of nanosized silicon-based slurries create difficulties in anode production, such as uneven coating and poor adhesion to current collectors. Optimizing the delicate interplay between particle size, porosity, cost, and manufacturability while maintaining high electrochemical performance has proven to be a significant challenge in the development of silicon anodes.

To overcome these challenges, we propose an approach that leverages plasma-produced extra-small silicon particles (x-Si, <10 nm) with a narrow size distribution. These particles are assembled into micron-sized structures with precisely controlled porosity. Figure 1a outlines the three main steps of our process: (1) plasma synthesis of x-Si particles, (2) evaporation-induced assembly of primary x-Si particles into micron-sized structures, and (3) carbon coating via chemical vapor deposition (CVD). This approach combines the advantages of nanostructured silicon with the scalability and ease of processing of micron-scale particles, offering a simple and scalable solution for silicon anodes. By carefully assembling the <10 nm silicon particles and encasing them in a carbon matrix, we have realized silicon-dominant anodes that retain 76% of their capacity after 500 cycles in pouch cells paired with $LiNi_{0.8}Mn_{0.1}Co_{0.1}O_2$ (NMC811) cathodes. This performance is achieved without pre-lithiation or pre-cycling because of the low initial lithium-ion inventory losses and the rapid stabilization of the SEI layer, with Coulombic efficiency (CE) climbing above 99.9% within the first 5 charge-discharge cycles. This study underscores the importance of controlling not only the size of the silicon particles but also the length scale and dispersion of porosity. It suggests that this is achievable by using small building blocks (i.e. <10 nm silicon particles) in combination with the appropriate assembly step.

2. **Results and Discussion**

The first step utilizes a radio frequency (RF) plasma to nucleate and grow the x-Si particles from silane ($SiH_4$) precursor.[12] This process is characterized by an inherent non-thermal equilibrium between free electrons and the background gas, resulting in rapid conversion of silane into solid particles with sub-10 millisecond reaction time and near 100% silane utilization.[12] The plasma electrostatically stabilizes the primary particles in the gas phase, slowing down aggregation, coalescence, and growth.[13] As a result, the x-Si particles are small (<10 nm) and exhibit a narrow size distribution (mean size 5.8 ± 1.1 nm), as shown in Figure S1a. The lognormal fits of the size distributions show a geometric standard deviation ($\sigma_g$) of 1.21, significantly narrower than the coagulation-dominated, self-preserving size distributions ($\sigma_g$ = 1.44) typically observed in aerosolized particles.[14] In contrast, commercially available silicon nanoparticles (obtained from Nanostructured & Amorphous Materials) are much larger (mean size 96 ± 60 nm) and exhibit a substantially broader distribution ($\sigma_g$ = 1.91, Figure S1b). The tight particle size control achieved through plasma synthesis inherently avoids the formation of particles larger than 150 nm, which are known to contribute to cracking and mechanical failure during lithiation.[6]

The plasma synthesis approach has already been successfully employed for a broad range of applications, including to produce silicon quantum dots and energetic materials.[15–17] Recently, the Neale group has published multiple reports on plasma-produced silicon particles for lithium-ion battery anodes, confirming their excellent cycle life. In these reports, the silicon surface is carefully engineered with molecular groups to passivate the surface and provide good electronic conductivity.[18–21] Our previous work on plasma-produced silicon quantum dots embedded in carbon matrices derived from annealed polymers also demonstrated highly stable anodes.[22,23] However, these anodes typically suffer from low initial Coulombic efficiency (ICE) because of their high surface area. We have found that a simple evaporation-driven assembly step solves this issue. This second step, shown in Figure 1a, utilizes the x-Si particles as building blocks to create compact micron-scale structures. Silicon particles are first sonicated in chloroform, which is then evaporated so that capillary forces form densely packed agglomerates. In the third and final step, the assembled particles (x-Si-A) are coated with carbon via CVD using acetylene ($C_2H_2$) as the precursor. Acetylene is first thermally cracked at a relatively moderate temperature (590°C), at a pressure of ~1.05 atm, to ensure the carbon infiltration into the porous structure. The particles are then annealed at 650-750°C in argon to increase the degree of graphitization of the carbon coating. This approach enhances the quality of the carbon shell, improving its electrical conductivity and the performance of the resulting anode.[24]

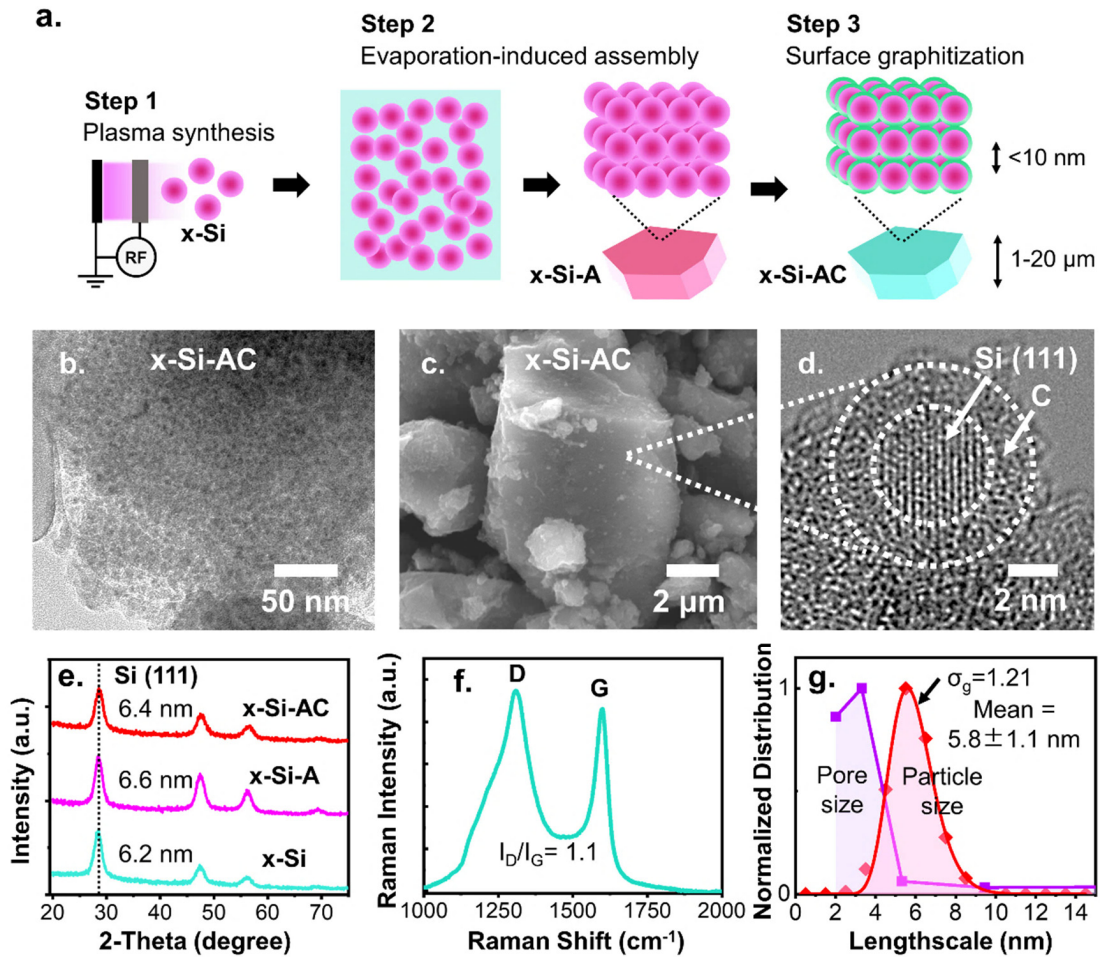

**Figure 1.** Synthesis and characterization of x-Si-AC particles. (a) Schematic of the three-step process: plasma synthesis of <10 nm x-Si particles, evaporation-induced assembly, and carbon coating via CVD. (b-d) Microscopy images showing the hierarchical structure of x-Si-AC particles: (b) TEM image of compact ensemble of <10 nm particles inside a microparticle, (c) SEM image of micron-sized particles, and (d) high-resolution TEM image of a single silicon particle with a graphitized carbon shell. (e) XRD patterns illustrating structural evolution during synthesis stages. (f) Raman spectrum of the final x-Si-AC composite. (g) Size distribution of silicon particles and pores in x-Si-AC, highlighting the sub-10 nm length scales for both silicon particles and pores.

The final material, x-Si-AC, comprises micron-sized particles resembling graphite flakes, composed of tightly packed sub-10 nm silicon particles encased in a graphitized carbon matrix (Figure 1b-d). The size of the assembled microparticles ranges from ~0.6-8 μm, with an average size of ~2.2 μm (Figure S2). Transmission electron microscopy (TEM) images in Figures 1b and S3b reveal a compact ensemble of silicon nanocrystals within the microparticle, forming an open and porous aggregates (Figure S3a). High-resolution TEM of an individual ~5 nm silicon particle shows distinct Si (111) planes surrounded by a thin graphitized carbon shell (Figure 1d).

X-ray diffraction (XRD) patterns shown in Figure 1e illustrate the evolution of the material structure during the synthesis, assembly, and carbon coating stages. Scherrer analysis indicates a particle size of ~6 nm, in agreement with the TEM analysis (Figures S1a and 1g). This small crystallite size is maintained throughout the assembly and CVD steps, as shown in Figure 1e. Annealing the assembled particles for 1 hour in argon at the same CVD temperature leads to particle growth to ~10 nm, based on XRD analysis. Raman spectroscopy of the final x-Si-AC composite (Figure 1f) confirms the nanoscale graphitic nature of the carbon coating. The spectrum exhibits characteristic D (~1300 cm$^{-1}$) and G (~1600 cm$^{-1}$) peaks with an $I_D/I_G$ ratio of ~1.1. The G peak position at a higher frequency than typically observed in purely amorphous carbon (~1500 cm$^{-1}$) indicates the presence of sp²-rich graphitic domains.[25] Concurrently, the relatively broad D peak and $I_D/I_G$ ratio suggest some sp³-bond disorder, likely due to the high curvature and edges of the sub-10 nm particles. These observations, combined with the partially disordered graphitic planes observed in high-resolution TEM (Figure 1d), indicate a turbostratic carbon structure containing locally ordered, sp$^2$-domains interspersed with defects. Most importantly, Figure 1g underscores the defining feature of x-Si-AC microparticles: silicon particle and pore length scales are both below 10 nm. This hierarchical structure, integrating nanoscale features within micro-sized assemblies, enables superior electrochemical performance, as discussed in the next sections.

The morphology and elemental composition of the particle assemblies were investigated using high-angle annular dark-field scanning transmission electron microscopy (HAADF-STEM), along with energy dispersive spectroscopy (EDS). Figure 2a shows the HAADF-STEM images of x-Si particles with and without assembly (x-Si-AC and x-Si-C, respectively). Both configurations exhibit sub-10 nm silicon particles coated with carbon, but their morphologies are markedly different. Without evaporative assembly, the particles form highly porous aggregates with an open-network fractal-like morphology, characteristic of aerosol processes where nucleation is followed by rapid, irreversible sticking, leading to random diffusion-limited assemblies.[26] EDS mapping in Figure 2a confirms uniform carbon infilling of the assembled silicon particles via the CVD procedure. SEM-EDS analysis of larger samples indicates a silicon content of approximately 54-60% by weight.

The pore structure of the different particle configurations was further quantitatively probed through nitrogen sorption porosimetry and tap density measurements. Table 1 summarizes the porous properties of all samples, with specific surface area determined via Brunauer-Emmett-Teller (BET) analysis and pore size distributions via Barrett-Joyner-Halenda (BJH) analysis. The tap densities, which reflect bulk packing efficiency, are also listed. The pore size distributions for all silicon samples explored in this study are presented in Figure S4 in the Supporting Information. As shown in Table 1, the as-synthesized (i.e., without assembly) silicon particles (x-Si) exhibit a large average pore size of 11.2 nm, a substantial pore volume of ~0.95 cm$^3$ g$^{-1}$, a wide range of pore sizes (5->50 nm), and a very low tap density (0.04 g cm$^{-3}$), consistent with the highly porous structure observed in the TEM images (Figure S3a). Applying a carbon coating to these particles without any assembly (x-Si-C) results in partial reduction in pore size and volume but does not significantly change the tap density. In contrast, assembling the x-Si particles (x-Si-A) dramatically affects pore characteristics, reducing the average pore size from 11.2 nm to 3.0 nm, decreasing the pore volume by ~52% (from 0.95 cm$^3$ g$^{-1}$ to 0.45 cm$^3$ g$^{-1}$), and increasing the tap density by more than an order of magnitude (from 0.04 g cm$^{-3}$ to 0.59 g cm$^{-3}$). Carbon coating the assembled particles (x-Si-AC) further increases the tap density to ~1.2 g cm$^{-3}$ while maintaining the sub-10 nm pore size (2.9 nm). The tap density of the assembled, carbon coated silicon powders is higher than commercial graphite powders. The pore volume in x-Si-AC particles is reduced to 0.06 cm$^3$ g$^{-1}$, representing a ~16-fold densification compared to the starting x-Si particles. These quantitative results are visually corroborated by the STEM-EDS micrographs in Figure 2a, which show very densely packed constituents in the x-Si-AC assemblies.

| Material | Specific surface area (m$^2$ g$^{-1}$) | Average pore size (nm) | Pore volume (cm$^3$ g$^{-1}$) | Tap density (g cm$^{-3}$) |
|---|---|---|---|---|
| x-Si | 351 | 11.2 | 0.95 | 0.04 |
| x-Si-A | 426 | 3.0 | 0.45 | 0.59 |
| x-Si-AC | 84 | 2.9 | 0.06 | 1.22 |
| x-Si-C | 307 | 7.7 | 0.53 | 0.05 |
| Com-Si-AC | 21 | 9.2 | 0.04 | 0.26 |

**Table 1.** Pore structure characteristics and tap densities of different silicon particle-pore configurations. x-Si: as produced silicon particles; x-Si-A: silicon particles after evaporative assembly; x-Si-AC: assembled silicon particles after CVD coating with carbon; x-Si-C: CVD coated silicon particles, without assembly step; Com-Si-AC: commercial silicon particles, assembled and CVD-coated with carbon.

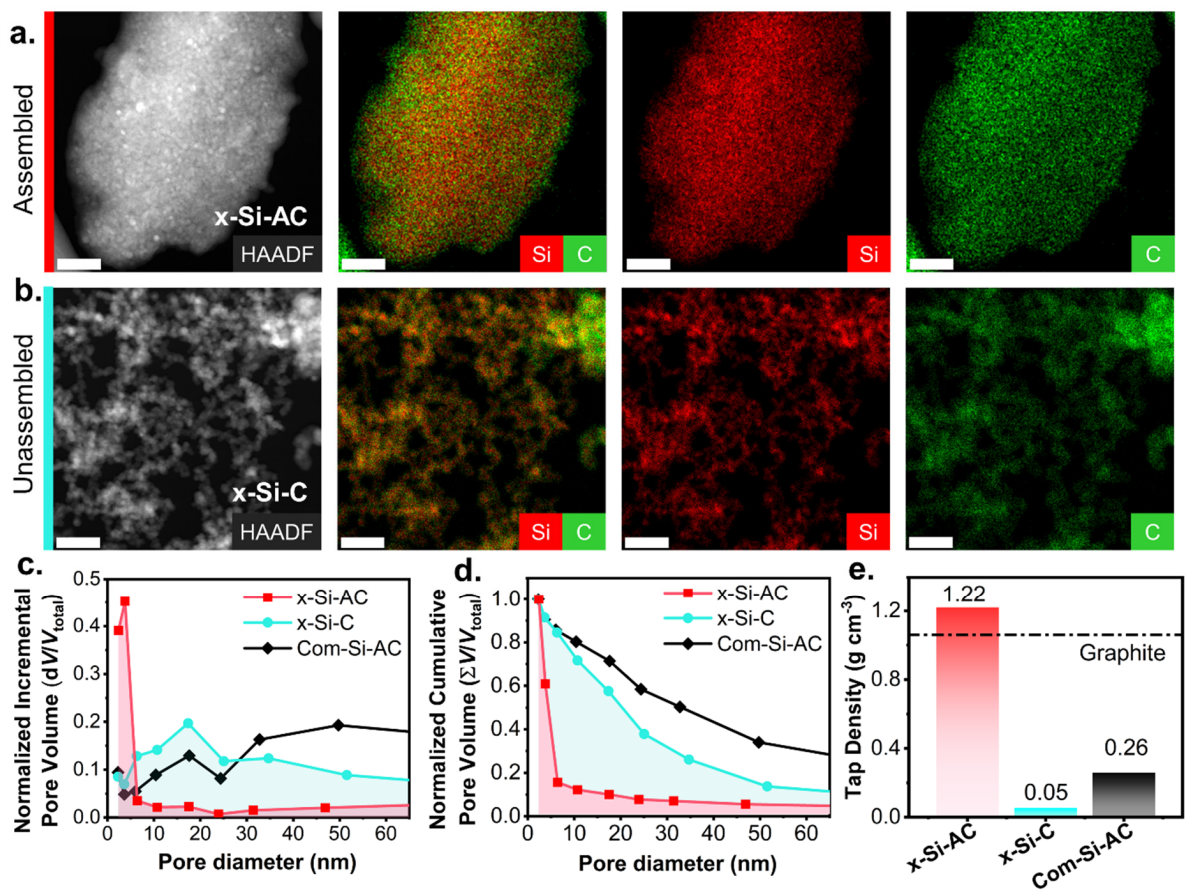

**Figure 2.** Morphology and pore structure characterization of silicon particle-pore assemblies. (a) HAADF-STEM images and EDS maps of assembled (x-Si-AC) and unassembled (x-Si-C) <10 nm silicon particles. (b) Normalized incremental and (c) cumulative pore volume distributions for x-Si-AC, x-Si-C, and Com-Si-AC structures as obtained from BJH analysis of $N_2$-sorption porosimetry data. (d) Comparison of tap densities for the three assemblies. The assembled x-Si-AC particles exhibit a highly compact structure with narrowly distributed sub-10 nm pores and high tap density comparable to graphite. Scale bars: 50 nm.

To better understand the impact of our assembly process, we compared our particles to commercially available silicon particles obtained from Nanostructured and Amorphous Materials that underwent similar assembly and carbon coating steps (Com-Si-AC). The pore size distributions of x-Si-AC, x-Si-C, and Com-Si-AC samples are visually depicted in Figures 2b and 2c, which show the normalized incremental and cumulative pore volumes, respectively. As shown in Figure 2b-c, the x-Si-AC sample exhibits a narrow pore-size distribution, with approximately 90% of the total pore volume coming from sub-10 nm pores. In contrast, the unassembled as well as commercial silicon counterparts show much broader pore-size distributions, with only about 15% of the pore volume contributed by sub-10 nm pores. The commercial silicon-based assemblies (Com-Si-AC) demonstrate the effect of heterogeneous particle sizes on the resulting pore structure. These assemblies show much larger pore sizes (>50 nm) and a relatively low tap density of 0.26 g cm$^{-3}$. This can be attributed to their broad particle size distribution, which leads to inefficient packing and larger inter-particle pores, as evidenced in the TEM images shown in Figure S5. In contrast, the small size and narrow particle size distribution of the plasma-produced particles lends itself to a similarly narrow pore size distribution upon assembly, with both particles and pores predominantly in the sub-10 nm range (see Figure 1f).

The effective densification of the <10 nm particles using our approach is further highlighted by the 30-fold increase in tap density from 0.04 g cm$^{-3}$ to 1.22 g cm$^{-3}$ and a 4-fold reduction in the specific surface area from ~350 m$^2$ g$^{-1}$ to ~80 m$^2$ g$^{-1}$, as shown in Figure 2d. The tap density of the x-Si-AC particles exceeds that of commercial graphite particles, significantly enhancing their processability. The tap density of the x-Si-AC microparticles (1.22 g cm$^{-3}$) significantly surpasses that of similar evaporation-driven approaches using commercial silicon powders, such as the pomegranate structure (0.53 g cm$^{-3}$),[8] and is comparable to some of the highest tap densities reported for silicon powders obtained through high-pressure mechanical compaction methods (~1.38 g cm$^{-3}$).[27] To summarize, the carbon-coated, assembled x-Si particles (x-Si-AC), exhibit the highest tap density, low specific surface area and pore volumes, while exhibiting very narrowly distributed sub-10 nm pores. Their size distribution is small and narrow, both in terms of primary particle and pore sizes. The implications of this unique structure for electrochemical performance are discussed next.

To understand the interplay between pore and grain sizes of the assemblies on electrochemical characteristics, half-cell tests were conducted using lithium metal as the counter electrode. Silicon-dominant anodes were prepared by mixing 75 wt.% of the active material (carbon-coated, assembled x-Si-AC), 10 wt.% Super P carbon black, and 7.5 wt.% each of carboxymethylcellulose (CMC) and polyacrylic acid (PAA). The same composition was used for all samples to focus on the effects of the inherent particle-pore structure. Figure 3 presents the electrochemical data for the three configurations: x-Si-AC (small particles, small pores), x-Si-C (small particles, large pores), and Com-Si-AC (large particles, large pores). The capacities reported are calculated based on the total weight of the active Si-C material, which constitutes 75 wt.% of the total anode weight. As shown in Figure 3a, the x-Si-based anodes outperform their commercial counterparts, with both x-Si-AC and x-Si-C anodes exhibiting high cycling stability. The x-Si-AC anode material demonstrates a high first-cycle discharge capacity of 1540 mAh $g^{-1}$ and a charge capacity of 1300 mAh $g^{-1}$ at a 0.1C cycling rate. Considering that the x-Si-AC composites consist of ~54-60% silicon by weight (based on SEM-EDS), the silicon contribution to the first cycle discharge capacity is estimated to be ~2570-2850 mAh $g^{-1}$. This suggests ~72-80% utilization of the theoretical silicon capacity. Despite this high silicon utilization, the x-Si-AC anode exhibits high capacity retention (~94%) over 60 deep charge-discharge cycles (0.1 C rate). In contrast, the commercial silicon-based anode delivers a high initial specific capacity (~2200 mAh $g^{-1}$) but suffers from rapid capacity decay, dropping by a factor of 2 within 50 cycles. This is likely due to particle cracking issues caused by a high volumetric proportion of larger particles (>150 nm, Figure S1b).

The CE profiles shown in Figures 3b and 3c provide further insights into SEI formation and lithium inventory loss among the three samples. The x-Si-C (i.e. small silicon particles without assembly step) anode exhibits a very low initial CE of 51.8% and requires >20 cycles to reach a stabilized CE value of ~98.30%. Similarly, while the Com-Si-AC anode shows a relatively high initial CE of 92%, it also suffers from poor CE stabilization, reaching only ~98.75% (Figures 3b and 3c). The x-Si-AC anode, on the other hand, combines excellent cycling stability with fast CE stabilization. It achieves a high initial CE of 85% and stabilizes rapidly to >99% in the first 3 cycles, reaching ~99.91% within the first 5 cycles (Figure 3c). The CE stabilization differences can also be visualized by the cumulative efficiency data shown in Figure 3d. Cumulative efficiency, defined as the cumulative product of half-cell CE values, has been proposed as a valuable metric to predict full-cell stability from half-cell data.[28] For long-term stability in full-cells, the cumulative efficiency must be maximized, as lithium losses compound over each cycle and gradually deplete the limited lithium inventory. As shown in Figure 3d, both Com-Si-AC and x-Si-C anodes exhibit a rapid decline in cumulative efficiency. For commercial particles, this can be attributed to the expansion and cracking of larger particles, which expose new silicon surfaces to the electrolyte and prevent the formation of a stable SEI. In the case of x-Si-C, while cracking is mitigated due to smaller <10 nm particles, the excessive porosity and accessible surfaces lead to elevated reactivity with the electrolyte and continued lithium loss. As a result, after 60 cycles, the cumulative efficiency for these samples is highly diminished (~30% for Com-Si-AC and ~10% for x-Si-C). In stark contrast, the x-Si-AC anode, with its extremely small pores *and* particles, maintains a considerably higher cumulative efficiency of ~75% after 60 cycles, reflecting significantly reduced lithium loss due to its compact *and* stable structure (see Figure 2).

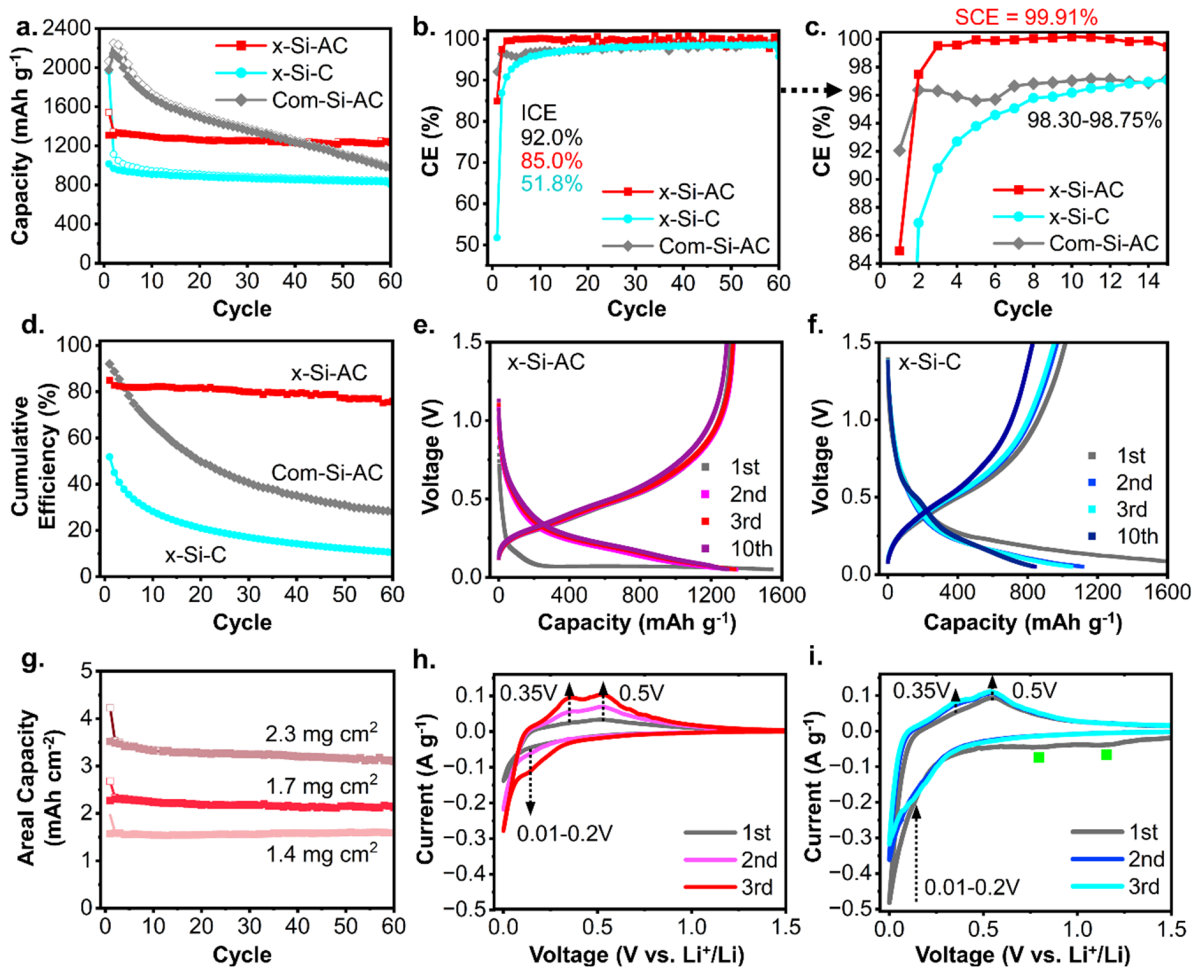

**Figure 3.** Electrochemical performance of silicon particle-pore assemblies (x-Si-AC, x-Si-C, and Com-Si-AC) in half-cells: (a) Capacity and (b) CE over charge-discharge cycles at C/10 rate. (c) Initial CE stabilization (enlarged view of b). (d) Cumulative efficiency over cycles. (e, f) Charge-discharge voltage profiles for x-Si-AC and x-Si-C. (g) Areal capacity versus mass loading for x-Si-AC particles. (h, i) CV of x-Si-AC and x-Si-C. The compact <10 nm particle-pore architecture of x-Si-AC enable superior electrochemical properties compared to other structures.

The charge-discharge voltage profiles for x-Si-AC and x-Si-C anodes at 0.1C rate are shown in Figures 3e and 3f, respectively. The lithiation and delithiation plateaus (~0.2-0.01V and ~0.3-0.5V, respectively) correspond to the alloying and dealloying reactions of silicon with lithium. After the first discharge cycle, the x-Si-AC anode exhibits rapid stabilization, as evidenced by the nearly identical voltage profiles from cycles 1, 2, 3, and 10. In contrast, the voltage profiles for x-Si-C show a progressive decline with each cycle. Interestingly, the x-Si-C anode exhibits a higher overpotential compared to x-Si-AC, suggesting slower lithiation kinetics due to low particle-particle contact and long lithium diffusion length scales in the porous aggregates. In other words, the compact structure of x-Si-AC may also exhibit enhanced lithiation kinetics due to shorter diffusion lengths, higher interparticle contact and conductivity. Cyclic voltammetry (CV) curves for x-Si-AC and x-Si-C anodes for the first three cycles at a scan rate of 0.02 mV s$^{-1}$ are shown in Figures 3h and 3i, respectively. Both exhibit characteristic lithiation and delithiation peaks for silicon in the ~0.01-0.2V and ~0.35-0.5V regions, respectively. The x-Si-C anode shows distinct irreversible peaks in the 0.8-1.1V region (marked in green in Figure 3i), corresponding to electrolyte reduction (ethylene carbonate at ~0.8V and fluoroethylene carbonate at ~1.1V).[29,30] These peaks are absent in the x-Si-AC anode, indicating suppressed reactivity with the electrolyte. Additionally, the x-Si-C anode shows sharp lithiation peaks in the first cycle (~0.01-0.2V) that diminish in subsequent cycles, reflecting irreversible losses. The delithiation peaks (~0.35V and ~0.5V) also do not increase with cycling. In contrast, the x-Si-AC anode demonstrates increasing lithiation and delithiation peaks over subsequent cycles, indicating higher reversibility and activation of silicon as the SEI stabilizes. Conclusively, these results suggest that the compact structure of x-Si-AC promotes stable SEI formation and enhanced electrochemical reversibility. Here, it is also important to emphasize that, without the assembly step, x-Si-C aggregates form highly viscous slurries that make coating quite challenging. In comparison, the compact nature and high tap densities of x-Si-AC microparticles enable the fabrication of thicker electrodes with higher areal capacities. This is demonstrated in Figure 3g, which shows that practical areal capacities of ~3 mAh cm$^{-2}$ can be easily achieved by simply coating thicker films of x-Si-AC while maintaining good electrochemical performance.

The assembled x-Si-AC anodes were also evaluated in single-layer pouch-type full cells using a Li(Ni$_{0.8}$Mn$_{0.1}$Co$_{0.1}$)O$_2$ (NMC811) and LiFePO$_4$ (LFP) cathodes, demonstrating compatibility with different cathode chemistries (Figure 4). These cells, constructed by SpectraPower LLC, were tested under standard conditions using the full voltage range of 2.8-4.2V for NMC811 and 2.0-3.6V for LFP cathodes. N/P ratios of ~1.06:1 (NMC811) and ~1.3:1 (LFP) were used, based on the stable reversible capacity after initial formation. A standard cycling protocol was employed, typically consisting of formation cycles at C/20 and C/5 followed by steady-state cycling at C/3. Specific details for each cell type are provided in the Materials and Methods section. These silicon-dominant anodes contain ~45-50% silicon by weight, with the remainder consisting of carbon coating, binders, and conductive additives, without any graphite dilution. Reversible areal capacities of these anodes (after first cycle losses) ranged from ~1.5-2 mAh cm$^{-2}$ in half cells, with a total anode reversible capacity of ~980-1220 mAh g$^{-1}$ (including all components). Figures 4a and 4c present the charge-discharge cycling data for the NMC811 and LFP pouch cells, respectively. In both cases, x-Si-AC anodes show excellent cycle life, retaining ~80% of the initial capacity for over 400 cycles. These results were achieved without pre-lithiation, pre-cycling, graphite blends, or special cycling protocols, confirming the potential of this material as a true drop-in replacement to graphite.

Figures 4b and 4d show the early formation data of the full cells. First cycle CE values range from ~67-78%. Following the first cycle, the CE rapidly increases to more than 99.5% by cycle 4 for NMC811 cells and 99.2% for LFP cells, stabilizing at ~99.9% for the remaining cycles in both cases. This data mirrors the fast stabilization observed in half-cells (Figure 3c) and reflects the formation of a stable SEI layer on the anode particles. For silicon-rich anodes, techniques like pre-lithiation or pre-cycling are typically used to mitigate initial lithium losses and extend cycle life in silicon-based full cells.[10,11,31] However, these methods add complexity to cell manufacturing and pose a significant barrier to practical implementation. The high cumulative efficiencies of x-Si-AC anodes, as shown in Figures 3d, enable the realization of stable silicon-dominant anodes without these additional steps, thus facilitating more straightforward scale-up and integration into commercial applications.

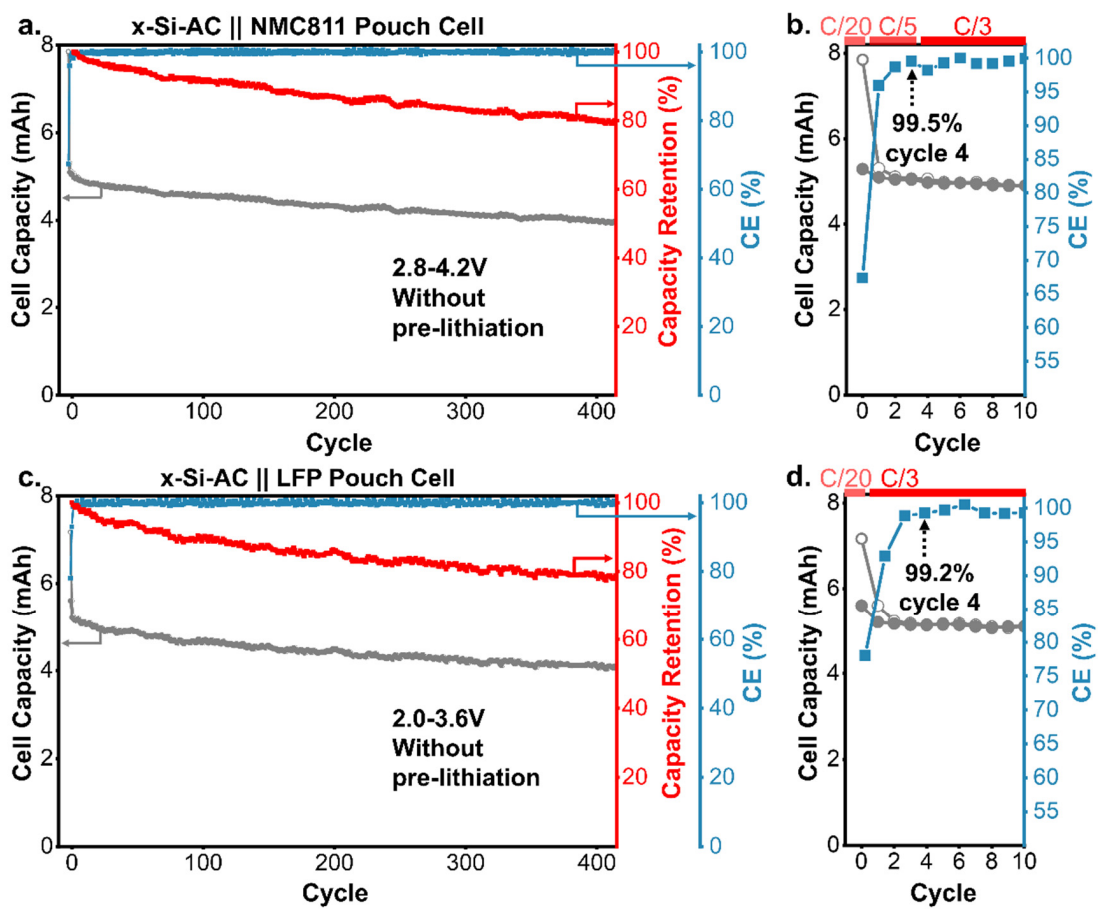

**Figure 4.** Electrochemical performance of x-Si-AC anodes in full pouch cells with different cathode chemistries. (a,b) x-Si-AC || NMC811: (a) Long-term cycling stability, demonstrating high capacity retention and near 100% Coulombic efficiency over 400 cycles without any pre-lithiation or pre-cycling. (b) Initial formation cycles, highlighting rapid CE stabilization. (c,d) x-Si-AC || LFP: (c) Long-term cycling performance and (d) initial formation cycles. These results place x-Si-AC among the most stable silicon-dominant anodes reported to date.

The key to the high stability of assembled x-Si-AC particles lies in the synergistic effect of <10 nm silicon particles and <10 nm porosity, as schematically depicted in Figure 5. This structure addresses two primary limitations associated with silicon anodes. First, as pointed out by Nava *et al.* and Schwan *et al.*,[24,32] commercially available silicon particles, even with an average size below 150 nm, inevitably contain a fraction of larger particles (see Figure S1b). These larger particles, even in small proportions, occupy a significant volume fraction (scaling as $d_P^3$, $d_P$ being particle size) and can lead to cracking and capacity fade. The plasma-produced x-Si particles used in this study ensure all particles are <10 nm (well below the critical size), enabling high cycle life (Figures 1g and S1a). Secondly, the x-Si-AC structure (with <10 nm particles *and* pores) mitigates the issue of unstable SEI formation common in nanostructured silicon anodes. Typically, high surface area combined with volume changes during cycling leads to excessive SEI growth and low initial and stabilized CE values. The unassembled <10 nm particles (x-Si-C) shown in Figure 5 represent an extreme case of this effect, where very high surface areas and highly porous structure result in very low CE values, excessive SEI growth, and significant lithium-inventory loss. The structure we have developed in this study (x-Si-AC) addresses this problem by volumetrically limiting SEI growth. This is because the pore size is smaller than the typical thickness of the SEI layer, which is several tens of nanometers on graphite[33,34] and has been reported below 50 nm for silicon anodes with FEC-containing electrolytes.[35,36] Post-cycling STEM-EDS analysis of x-Si-AC microparticles cycled in half-cells reveals a ~40-50 nm C/O-rich SEI layer predominantly on the outer surface of microparticles (Figure S6a). The thickness of the SEI layer exceeds the internal pore size (<5 nm) of the microparticles, effectively sealing the pores and preventing further electrolyte penetration, as illustrated in Figure 5. Furthermore, cycling x-Si-AC anodes in half-cells with and without fluoroethylene carbonate (FEC) yields negligible differences in CE stabilization (Figure S6b). This contrasts sharply with studies showing significantly lower stabilized CE values in the absence of FEC (~97% without FEC vs. ~99% with).[37] This finding further supports the argument that electrolyte penetration into the microstructure is limited, demonstrating that SEI stabilization on the x-Si-AC microparticles is primarily a function of the inherent structure rather than electrolyte-induced SEI stabilization. The combination of <10 nm particles and <10 nm pores in x-Si-AC enables high stability by limiting particle expansion and constraining SEI growth, allowing high CE values despite the high surface area.

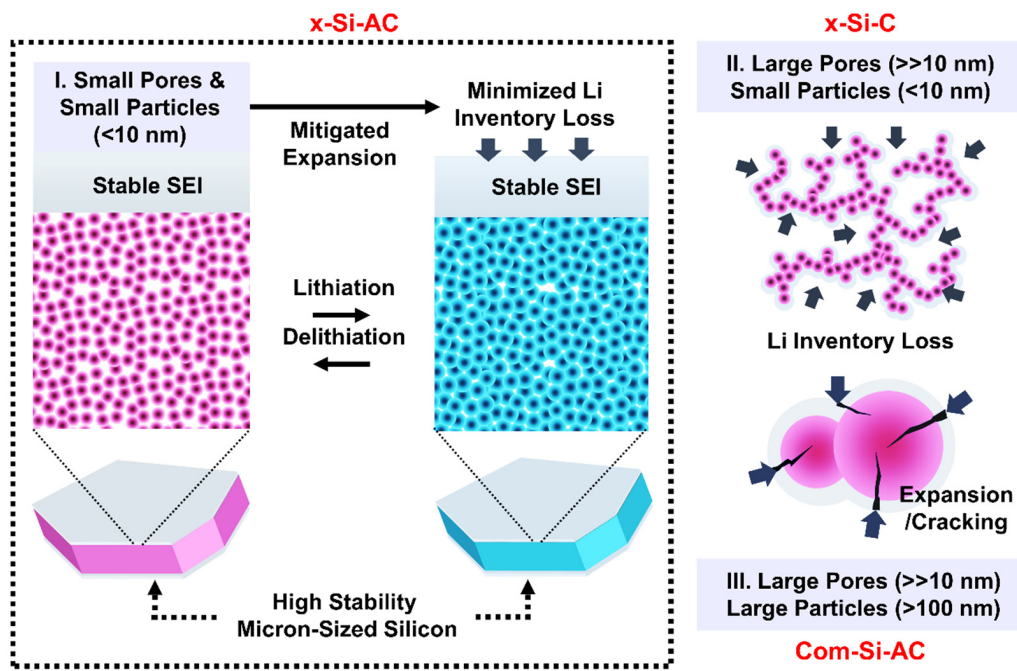

**Figure 5.** Schematic illustrating stability mechanisms in anodes based on different silicon particle-pore length scales. The assembled x-Si-AC microparticles (I) with <10 nm particles and pores demonstrate superior stability by synergistically limiting both particle expansion and SEI growth, compared to structures with larger pores (II) or larger particles and pores (III).

To summarize, this work demonstrates that highly stable silicon-dominant anodes are achievable through two key factors: (1) utilizing small silicon particles with a narrow size distribution as building blocks, and (2) achieving a uniform distribution of small pores within the assembled structure. When tested in pouch cells, this combination enables high Coulombic efficiency and excellent cycle life. The non-thermal plasma process employed here efficiently produces ultra-small particles on a millisecond timescale, readily achieving the necessary small particle sizes.[38] The evaporation-induced assembly process is simple, involving only sonication of the powder followed by solvent evaporation. The subsequent CVD process for carbon infiltration uses a commodity precursor and standard equipment. The simplicity of this procedure, coupled with the good performance of the resulting material, makes this technology promising for scale-up and large-scale implementation in next-generation silicon-containing lithium-ion batteries. This work highlights the need to precisely control not only silicon particle size but also porosity length scale. This is an additional important parameter that can be engineered to improve the characteristics of silicon-based anode materials.

3. **Materials and Methods**

*Synthesis of x-Si particles:* The extra-small silicon particles were produced using a large version of the flow-through plasma reactor first introduced by Mangolini et al.[38] A mixture of silane and argon (1.36% $SiH_4$ in Ar) was fed to a tubular quartz reactor with a two-inch diameter, with the reactor pressure in the range of 2.5-3 Torr. A radio-frequency electrode was used to sustain the plasma, with a typical power of 100W. The precursor was fed continuously to the reactor volume, and silicon particles were continuously nucleated, grown, and collected downstream of the plasma on a stainless-steel metal cloth. Analysis of the exhaust gas via FTIR showed that the plasma fully consumes the silane gas. For the two-inch reactor used here, roughly 50% of the incoming silicon mass is collected as particles downstream of the reactor. After particle synthesis, the reactor is evacuated and then filled with argon and the particles are transferred to a glovebox to minimize oxidation.

*Assembly and carbon coating:* The x-Si particles are then dispersed in a non-polar solvent (chloroform) at solid loadings of ~25-50 mg $mL^{-1}$ and ultrasonicated for ~1 hour. This results in a relatively stable and homogeneous dispersion of particles. The dispersion is then dried on a clean, smooth surface, where capillary forces driven by solvent evaporation induce particles to assemble into dense agglomerates, which are then gently crushed to give micron-sized flakes (see Figure S3b). Finally, the micron-sized particles are locked into place by a carbon layer applied via CVD. The assembled powders are placed in an alumina crucible enclosed in a two-inch diameter tubular quartz reactor. The reactor is evacuated and purged with argon, followed by the introduction of acetylene ($C_2H_2$) at a flow rate of ~200 sccm until a slightly positive pressure of ~1.05 atm is achieved. The sample is then heated to ~590°C at a heating rate of 20°C min$^{-1}$, followed by a 35 min hold at 590°C. After this, the reactor is evacuated, and argon is introduced again at a flow rate of 400 sccm to purge out any residual $C_2H_2$. The samples are then heated at 650-750°C for 10 minutes to improve the quality of the carbon coating.[24] The x-Si-C control samples (directly carbon-coated without assembly) were prepared by collecting the plasma-synthesized particles on a mesh filter placed inside the CVD quartz tube. Following particle synthesis, the system was flushed with argon. The CVD procedure was then carried out to coat the x-Si particles with carbon without exposing them to air. The same protocol was followed for commercially available silicon particles starting from the assembly step to obtain the Com-Si-AC composites.

*Material Characterization:* Particle morphology on the micron scale was analyzed with a Thermo Fisher Scientific NNS450 scanning electron microscope at a 15 kV accelerating voltage. An FEI Titan Themis 300 transmission electron microscope was used to perform STEM-HAADF, HRTEM, and elemental analysis on the particle assemblies on the nanoscale. Powder X-ray diffraction was obtained using a PANalytical Empyrean Series 2 with Cu Kα radiation. The Raman spectra of the samples were measured using a DXR SmartRaman microscope equipped with a 532 nm, 100 mW laser source. Measurements were taken within the spectral range of 1000 to 2000 cm$^{-1}$. A Micromeritics ASAP 2020 Plus physisorption instrument was used to obtain nitrogen-sorption porosimetry data (surface area, pore size, etc.) for different samples. Tap density measurements were performed by loading a known mass of powders into a 5 mL graduated cylinder, after which the cylinder was gently tapped onto a padded surface until no change in volume was observed (typically >500 times). A set of triplicates was performed for each sample, and an average was obtained.

*Electrode Preparation and Electrochemical Testing* All anodes explored in the study were composed of 75-83 wt% active material (x-Si-AC), 2-10 wt% conductive additives (Super P carbon black or carbon nanotubes), and 15 wt% of CMC and PAA with a 1:1 weight ratio. The specific compositional details are provided in the Supporting Information (Table S1). The active material and the carbon black were first crushed with a mortar and pestle, after which the appropriate water-based solutions of the binders (CMC-PAA) were added to prepare the slurry. The electrode slurries were then cast into electrodes on a copper current collector with typical loadings ranging from 1-3 mg cm$^{-2}$. The coated copper foils were then vacuum dried at ~90°C for 8 hours, followed by heating at 150°C for 2 hours. The second step induces cross-linking between the PAA and CMC components and improves the mechanical strength of the anode films.[39] Anode disks (12 mm) were then punched from the coated foils, which were then assembled into half cells against lithium metal (1 mm thickness) in a coin format (2032) in an argon glove box (O$_2$, H$_2$O<1 ppm).

The electrolyte used in half cells was a 1 M LiPF$_6$ solution in 1:1 v/v ethylene carbonate:diethyl carbonate (EC:DEC) with a 10 wt.% fluoroethylene carbonate (FEC) as an additive. Charge-discharge cycling measurements were performed on the coin cells between 0.05 and 1.5 V in a Neware Instruments battery tester with 0.1C as the deep-cycling current. Cyclic voltammetry (CV) analysis was performed on a Gamry Interface 1000 with a scan rate of 0.02 mV s$^{-1}$ from 2V to 0V. Single-layer pouch full cells were assembled and tested at an independent testing lab, SpectraPower LLC, using NMC811 and LFP as the cathodes. An N/P ratio of ~1.06 (NMC811) and ~1.3 (LFP) was used. 1.2 M LiPF$_6$ solution in 3:5 v/v ethylene carbonate:diethyl carbonate with a 10 wt.% FEC was used as the electrolyte for full cells. Cells were cycled at room temperature after build without any pre-lithiation or pre-cycling steps. For NMC811 cells, the formation cycles were performed sequentially with 1 cycle at C/20, 3 cycles at C/5, and continued cycling at C/3. For LFP cells, a simplified protocol of C/3 cycling was employed after the initial formation cycle at C/20. The full voltage range of 2.8-4.2V was used for the cells. A Neware BTS4000 Battery Testing System was used to test the cells. The compositional and electrochemical details of the anodes are summarized in the Supporting Information (Table S1).

ACKNOWLEDGEMENTS

This work has been primarily supported by the US Department of Energy, under award number DE-SC0024816. L.M. acknowledges the support of the US National Science Foundation, under award number 2053567, and the support of SiLi-ion Inc.

# Supporting Information

# Highly Stable Silicon Anodes Enabled by Sub-10 nm Pores and Particles

*Pankaj Ghildiyal[1]\*, Brandon Wagner[2], Jianjun Chen[3], Tu Nguyen[3], Aishwarya Belamkar[4], Juchen Guo[2,3], Lorenzo Mangolini[2,3]\**

1 SiLi-ion Inc., Multidisciplinary Research Building, 3401 Watkins Dr, Riverside, California 92507, United States

2 Materials Science and Engineering Program, University of California Riverside, Riverside, California 92521, United States

3 Department of Chemical and Environmental Engineering, University of California Riverside, Riverside, California 92521, United States

4 Department of Mechanical Engineering, University of California Riverside, Riverside, California 92521, United States

AUTHOR INFORMATION

**Corresponding Authors**

*Email: pankaj@siliion.com

*Email: lmangolini@engr.ucr.edu

**Supporting Figures and Tables**

**Particle size distributions for x-Si and commercial silicon particles:**

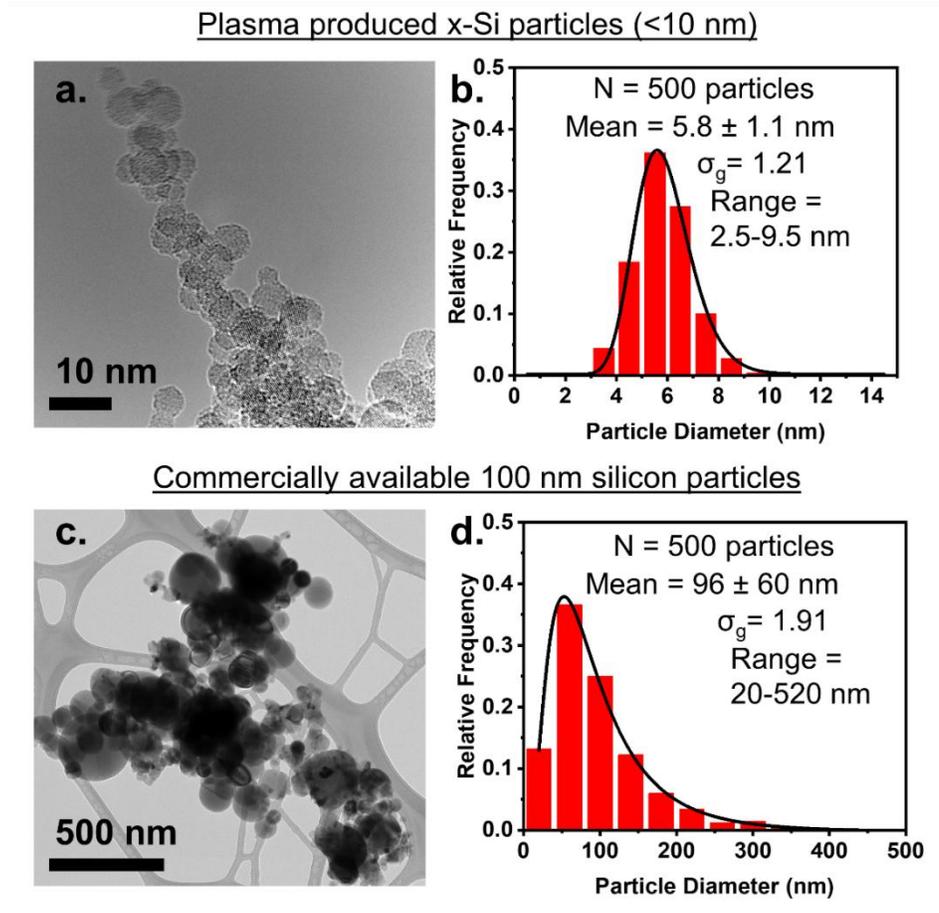

**Figure S1.** TEM images and corresponding lognormal particle size distributions of (a), (b) plasma-produced x-Si particles and (c), (d) commercially available 100 nm silicon particles from Nanostructured and Amorphous Materials. The x-Si particles exhibit a narrow size distribution with all particles below 10 nm, whereas the commercial particles show a broad size range of 20–500 nm. The geometric standard deviation () of the lognormal fit for the commercial particles is substantially higher than that of the x-Si particles. Particle size analysis was conducted using Nanomeasurer v. 1.2.5 on TEM images, with a total of 500 particles analyzed for both samples to generate the distributions.

**SEM of x-Si-AC microparticles prepared by assembling and carbon-coating x-Si particles:**

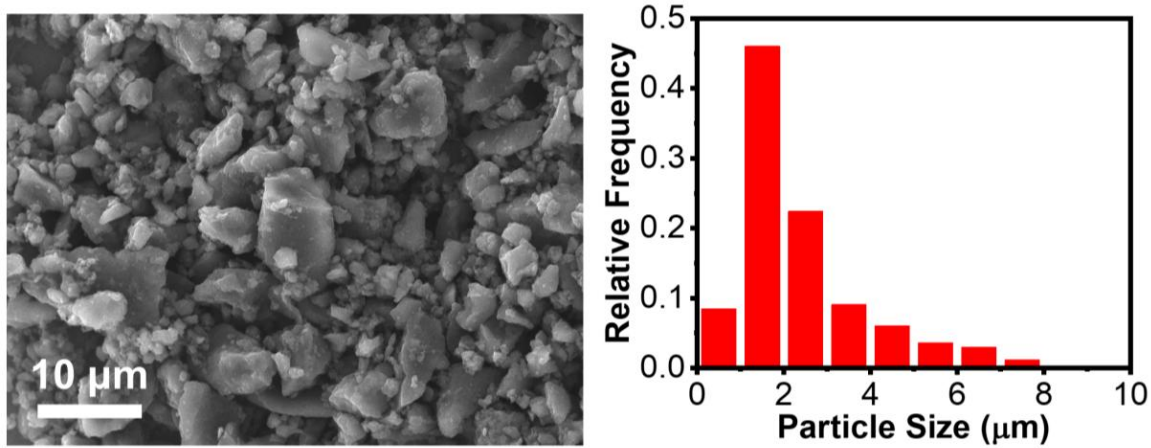

**Figure S2.** SEM image and corresponding particle size distribution dof x-Si-AC microparticles obtained after assembling and carbon-coating <10 nm Si particles. A total of 150 particles were considered for the size distribution.

**TEM of as-synthesized x-Si particles before and after assembly:**

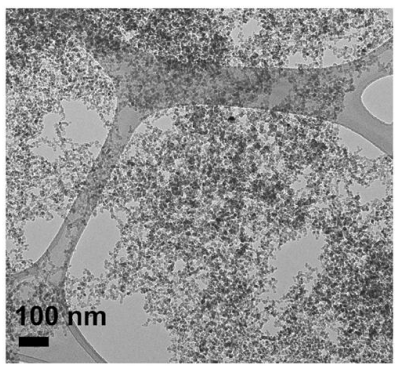 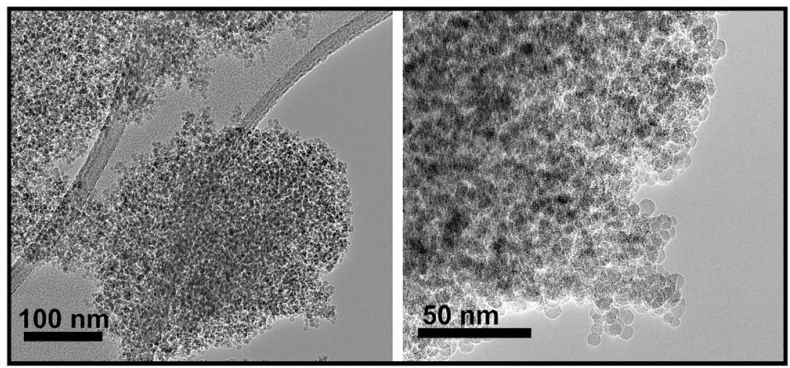

**Figure S3.** TEM images of <10 nm x-Si particles (a) before assembly (porous, open-network aggregates, and (b) after assembly (densely packed agglomerates).

**Pore size distributions of all silicon samples:**

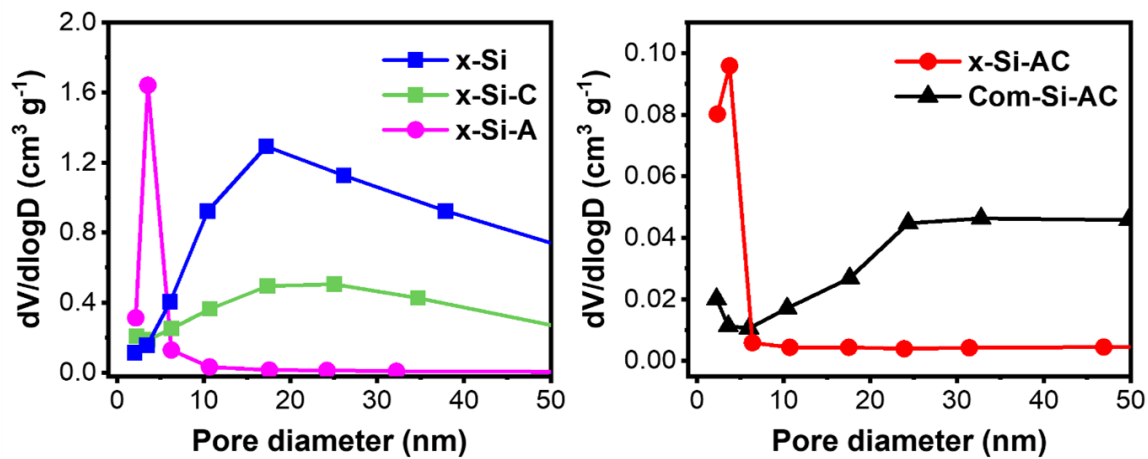

**Figure S4.** Pore size distributions obtained from $N_2$-desorption branch of the isotherm for all silicon samples explored in this study. The compact samples with low surface areas (x-Si-AC and Com-Si-AC) are shown in a separate graph (inset) due to their significantly lower incremental pore volumes.

**TEM of assembled commercial silicon particles after assembly and carbon coating:**

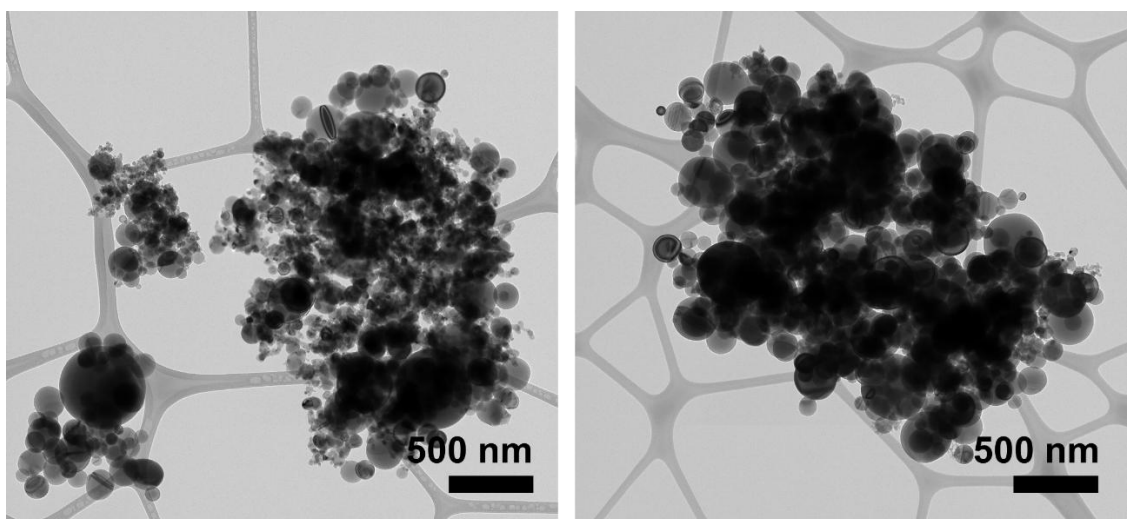

**Figure S5.** TEM images of commercial silicon nanoparticles after assembly and carbon coating (Com-Si-AC). The wide particle size distribution leads to inefficient packing and larger pore sizes.

**Post-cycling and electrolyte analysis of SEI formation in x-Si-AC microparticles:**

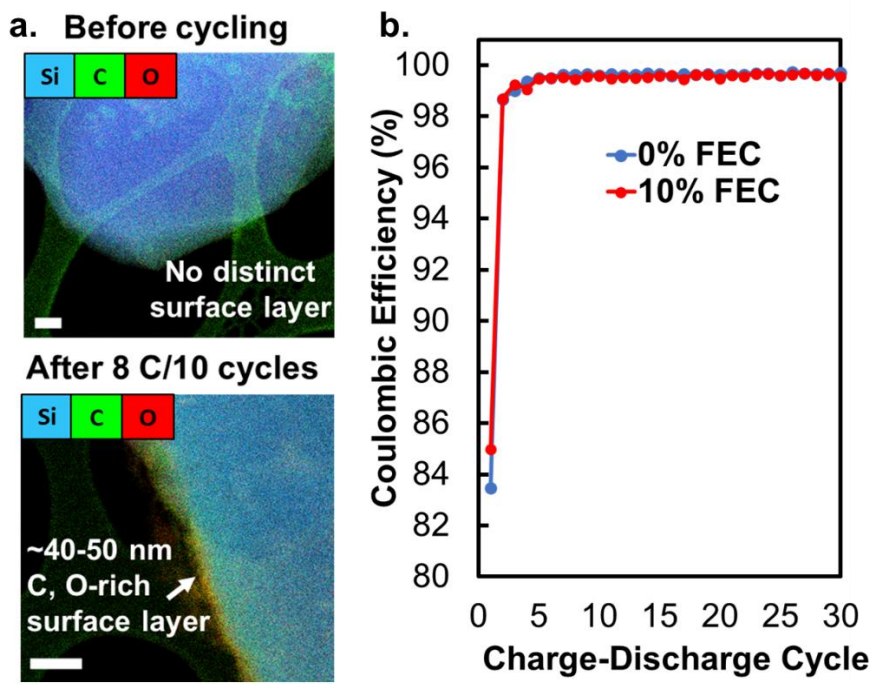

**Figure S6.** (a) STEM-EDS maps showing a 40-50 nm C/O-rich SEI layer primarily on the outer surface of cycled x-Si-AC microparticles, suggesting limited electrolyte penetration due to ultra-small (<5 nm) pores and compact structure. (b) CE stabilization of x-Si-AC anodes in half-cells with and without FEC additive, demonstrating effective SEI stabilization independent of FEC. Scale bars: 200 nm.

**Compositional and electrochemical details of different x-Si-based anodes explored:**

| Anode | Composition (wt. %) | Si content in active material (wt. %) | Si content in anode (wt. %) | 1$^{st}$-cycle total anode discharge capacity (mAh g$^{-1}$) | 1$^{st}$-cycle total anode discharge capacity (mAh g$^{-1}$) | ICE (%) |
|---|---|---|---|---|---|---|
| x-Si-AC | 75% active material, 10% carbon black, 7.5% PAA, 7.5% CMC | 55-60 | 41-45 | 1157 | 982 | 84.9 |
| x-Si-C | 75% active material, 10% carbon black, 7.5% PAA, 7.5% CMC | 32 | 24 | 1472 | 762 | 51.8 |
| Com-Si-AC | 75% active material, 10% carbon black, 7.5% PAA, 7.5% CMC | 88 | 66 | 1812 | 1668 | 92.0 |
| x-Si-AC (NMC811) | 75% active material, 10% carbon black, 7.5% PAA, 7.5% CMC | 55-60 | 41-45 | 1345 | 1117 | 83.1 |
| x-Si-AC (LFP) | 83% active material, 2% carbon nanotube additive, 7.5% PAA, 7.5% CMC | 55-60 | 46-50 | 1434 | 1215 | 84.7 |

**Table S1.** Composition (obtained from SEM-EDS) and electrochemical properties (obtained from half-cell data) of different silicon-based anodes explored in this study. The total Si content in the x-Si-AC-based anodes, including all components, ranged from ~41-50% by weight, with total reversible anode capacities ranging from ~982-1215 mAh g$^{-1}$.